\documentclass[aps,preprint,tightenlines,superscriptaddress,showpacs]{revtex4}
\usepackage{graphicx} 
\usepackage{dcolumn}  

\usepackage{epsfig}
\usepackage{refmerge}

\def \vec #1{\mbox{{\boldmath $#1$}}}

\def \GeV {{\rm GeV}}

\def \simgt {\stackrel{>}{\sim}}

\begin{document}


\preprint{\vbox{ \hbox{   }
                 \hbox{  }
                 \hbox{BELLE-CONF-0813}
                 \hbox{  }
}}                 

\title{ \quad\\[0.5cm]  High-statistics study of neutral-pion pair production\\
in two-photon collisions }

\affiliation{Budker Institute of Nuclear Physics, Novosibirsk}
\affiliation{Chiba University, Chiba}
\affiliation{University of Cincinnati, Cincinnati, Ohio 45221}
\affiliation{Department of Physics, Fu Jen Catholic University, Taipei}
\affiliation{Justus-Liebig-Universit\"at Gie\ss{}en, Gie\ss{}en}
\affiliation{The Graduate University for Advanced Studies, Hayama}
\affiliation{Gyeongsang National University, Chinju}
\affiliation{Hanyang University, Seoul}
\affiliation{University of Hawaii, Honolulu, Hawaii 96822}
\affiliation{High Energy Accelerator Research Organization (KEK), Tsukuba}
\affiliation{Hiroshima Institute of Technology, Hiroshima}
\affiliation{University of Illinois at Urbana-Champaign, Urbana, Illinois 61801}
\affiliation{Institute of High Energy Physics, Chinese Academy of Sciences, Beijing}
\affiliation{Institute of High Energy Physics, Vienna}
\affiliation{Institute of High Energy Physics, Protvino}
\affiliation{Institute for Theoretical and Experimental Physics, Moscow}
\affiliation{J. Stefan Institute, Ljubljana}
\affiliation{Kanagawa University, Yokohama}
\affiliation{Korea University, Seoul}
\affiliation{Kyoto University, Kyoto}
\affiliation{Kyungpook National University, Taegu}
\affiliation{\'Ecole Polytechnique F\'ed\'erale de Lausanne (EPFL), Lausanne}
\affiliation{Faculty of Mathematics and Physics, University of Ljubljana, Ljubljana}
\affiliation{University of Maribor, Maribor}
\affiliation{University of Melbourne, School of Physics, Victoria 3010}
\affiliation{Nagoya University, Nagoya}
\affiliation{Nara Women's University, Nara}
\affiliation{National Central University, Chung-li}
\affiliation{National United University, Miao Li}
\affiliation{Department of Physics, National Taiwan University, Taipei}
\affiliation{H. Niewodniczanski Institute of Nuclear Physics, Krakow}
\affiliation{Nippon Dental University, Niigata}
\affiliation{Niigata University, Niigata}
\affiliation{University of Nova Gorica, Nova Gorica}
\affiliation{Osaka City University, Osaka}
\affiliation{Osaka University, Osaka}
\affiliation{Panjab University, Chandigarh}
\affiliation{Peking University, Beijing}
\affiliation{Princeton University, Princeton, New Jersey 08544}
\affiliation{RIKEN BNL Research Center, Upton, New York 11973}
\affiliation{Saga University, Saga}
\affiliation{University of Science and Technology of China, Hefei}
\affiliation{Seoul National University, Seoul}
\affiliation{Shinshu University, Nagano}
\affiliation{Sungkyunkwan University, Suwon}
\affiliation{University of Sydney, Sydney, New South Wales}
\affiliation{Tata Institute of Fundamental Research, Mumbai}
\affiliation{Toho University, Funabashi}
\affiliation{Tohoku Gakuin University, Tagajo}
\affiliation{Tohoku University, Sendai}
\affiliation{Department of Physics, University of Tokyo, Tokyo}
\affiliation{Tokyo Institute of Technology, Tokyo}
\affiliation{Tokyo Metropolitan University, Tokyo}
\affiliation{Tokyo University of Agriculture and Technology, Tokyo}
\affiliation{Toyama National College of Maritime Technology, Toyama}
\affiliation{Virginia Polytechnic Institute and State University, Blacksburg, Virginia 24061}
\affiliation{Yonsei University, Seoul}
  \author{I.~Adachi}\affiliation{High Energy Accelerator Research Organization (KEK), Tsukuba} 
  \author{H.~Aihara}\affiliation{Department of Physics, University of Tokyo, Tokyo} 
  \author{K.~Arinstein}\affiliation{Budker Institute of Nuclear Physics, Novosibirsk} 
  \author{T.~Aso}\affiliation{Toyama National College of Maritime Technology, Toyama} 
  \author{V.~Aulchenko}\affiliation{Budker Institute of Nuclear Physics, Novosibirsk} 
  \author{T.~Aushev}\affiliation{\'Ecole Polytechnique F\'ed\'erale de Lausanne (EPFL), Lausanne}\affiliation{Institute for Theoretical and Experimental Physics, Moscow} 
  \author{T.~Aziz}\affiliation{Tata Institute of Fundamental Research, Mumbai} 
  \author{S.~Bahinipati}\affiliation{University of Cincinnati, Cincinnati, Ohio 45221} 
  \author{A.~M.~Bakich}\affiliation{University of Sydney, Sydney, New South Wales} 
  \author{V.~Balagura}\affiliation{Institute for Theoretical and Experimental Physics, Moscow} 
% ...
\collaboration{The Belle Collaboration}

\begin{abstract}
The differential cross section for the process $\gamma \gamma \to \pi^0 \pi^0$
has been measured in the kinematic range
0.6~GeV $< W < 4.1$~GeV, $|\cos \theta^*|<0.8$
in energy and pion scattering angle, respectively,
in the $\gamma\gamma$ center-of-mass system.
The results are based on a 223~fb$^{-1}$ data sample
collected with the Belle detector at the KEKB $e^+ e^-$ collider.
Using the data with $W > 1.4$~GeV, we obtain results on
light-quark resonances and charmonia. We also compare the observed
angular dependence and ratios of cross sections for neutral-pair
and charged-pair production to QCD models.
Differential cross sections are fitted in the energy region, 
$1.4~\GeV < W <  2.2~\GeV$, with a simple model where partial waves consist of 
resonances such as $f_2'(1525)$, $f_2(1950)$ and $f_4(2050)$ 
and smooth backgrounds.
In the higher energy region,
we observe production of the  $\chi_{c0}$ charmonium state and obtain the
product of its two-photon decay width and the branching fraction to 
$\pi^0\pi^0$.
The energy and angular dependences above 3.1~GeV are compatible 
with those measured in the $\pi^+\pi^-$ channel, and in addition we find
that the cross section ratio,
$\sigma(\pi^0\pi^0)/\sigma(\pi^+\pi^-)$, is $0.32 \pm 0.03 \pm 0.05$
on average in the 3.1-4.1~GeV region. 
\end{abstract}

\maketitle
\tighten
\normalsize
{\renewcommand{\thefootnote}{\fnsymbol{footnote}}
\setcounter{footnote}{0}


\section{Introduction}
Measurements of exclusive hadronic final states in two-photon
collisions provide valuable information concerning physics of light and 
heavy-quark resonances, perturbative and non-perturbative QCD 
and hadron-production mechanisms.
So far, we have measured
the production cross sections for charged-pion 
pairs~\cite{mori,nkzw}, 
charged- and neutral-kaon pairs~\cite{nkzw,wtchen}, 
and proton-antiproton pairs~\cite{kuo}.
We have also analyzed $D$-meson-pair production and observe a new
charmonium state~\cite{z3930}.
Recently, we have presented a measurement of neutral-pion pair
production based on a data sample corresponding to
an integrated luminosity of 95~fb$^{-1}$~\cite{pi0pi0}.
We have carried out an analysis to extract information
on light quark resonances from the energy and angular
dependences of the differential cross sections (DCS), by
fitting to the resonance parameters of $f_0(980)$, $f_2(1270)$ 
and other hypothetical resonances.

Here we present a measurement of the DCS, 
$d\sigma/d|\cos \theta^*|$, for the process $\gamma \gamma \to \pi^0 \pi^0$
in a wide two-photon center-of-mass (c.m.) 
energy ($W$) range from 0.6 to 4.1~GeV,
and in the c.m. angular range, $|\cos \theta^*| <0.8$.
We use a 223~fb$^{-1}$ data sample, which is more than twice 
as large as that in our previous analysis~\cite{pi0pi0, pi0pi0cp}.
We focus on the range $W > 1.4~\GeV$,
where the previous data was statistically limited.

In the intermediate energy range ($1.0~\GeV < W < 2.4~\GeV$), the formation
of meson resonances decaying to $\pi\pi$ is the dominant contribution.
For ordinary $q\bar{q}$ mesons in isospin conserving decays
to $\pi\pi$, the only allowed  $I^GJ^{PC}$ states
produced by two photons are  
$0^+$(even)$^{++}$, that is, $f_{J={\rm even}}$ mesons.
Several mesons with these quantum numbers are suggested by results of 
hadron-beam or charmonium decay experiments in the
1.5 - 2.2~GeV region, but
none of them have been firmly established in two-photon processes,
which are sensitive to the internal quark structure of the meson. 
In addition, the $\pi^0\pi^0$ channel has two advantages in the study of
resonances: a smaller contribution from the 
continuum is expected in it than in the $\pi^+\pi^-$ channel;
and the angular coverage is larger ($|\cos \theta^*| < 0.8$ 
instead of 0.6).

At higher energies, we can invoke a quark model.
In leading order calculations~\cite{bl,bc}, which take into account
the spin correlation between quarks, the $\pi^0\pi^0$
cross section is predicted to be much smaller
than that of $\pi^+\pi^-$, and the ratio 
of $\pi^0\pi^0$ to $\pi^+\pi^-$ cross sections is around 
0.03-0.06. 
However, higher-order or
non-perturbative QCD effects can modify this ratio. 
For example, the handbag model, which considers soft hadron exchange,
predicts the same amplitude for the two processes,
and thus the expected ratio is 0.5~\cite{handbag}. 
Analyses of energy and angular distributions of the cross sections
are essential for determining properties of the observed
resonances and for testing the validity of QCD models.

The organization of this article is as follows.
In Section~\ref{sec:appar}, a brief description of the Belle
detector is given.
Section~\ref{sec:cross} explains the procedure used to obtain differential
cross sections.
Section~\ref{sec:reson} is devoted to results on
resonances obtained by fitting differential cross sections 
in the range $1.4~\GeV < W < 2.2~\GeV$.
Section~\ref{sec:highe} describes analyses at higher energy.
The topics included there are the angular dependence as a function of $W$, 
the charmonia $\chi_{c0}$ and $\chi_{c2}$ states
and the ratio of $\pi^0 \pi^0$ to $\pi^+ \pi^-$ cross sections.
Finally, Section~\ref{sec:concl} concludes this report.
All of the results presented in this report are preliminary.

\section{Experimental apparatus}
\label{sec:appar}
We use a 223~fb$^{-1}$ data sample from the Belle experiment~\cite{belle}
at the KEKB accelerator~\cite{kekb}.
The data were recorded at $e^+e^-$ c.m. 
energies of 10.58~GeV (179~fb$^{-1}$),
10.52~GeV (19~fb$^{-1}$), 10.36~GeV ($\Upsilon(3S)$ runs,
2.9~fb$^{-1}$), 10.30~GeV (0.3~fb$^{-1}$) and 10.86~GeV
($\Upsilon(5S)$ runs, 21.7~fb$^{-1}$). 
The difference of two-photon fluxes per $e^+e^-$-beam luminosity 
($=$luminosity function) 
in the measured $W$ regions due to the difference of
the beam energies is small (maximum $\pm$ 4\%). 
We combine the results from the different beam energies. 
The effect on the cross section is less than 0.5\%.

The analysis is carried out in the ``zero-tag'' mode, where
neither the recoil electron nor positron are detected. 
We restrict the virtuality of the incident photons to be small
by imposing strict transverse-momentum balance with respect to 
the beam axis for the final-state hadronic system.

A comprehensive description of the Belle detector is
given elsewhere~\cite{belle}. 
We mention here only those
detector components that are essential for the present measurement.
Charged tracks are reconstructed from hit information in 
a silicon vertex detector and a central
drift chamber (CDC) located in a uniform 1.5~T solenoidal magnetic field.
The detector solenoid is oriented along the $z$ axis, which points
in the direction opposite to that of the positron beam. 
Photon detection and
energy measurements are performed with a CsI(Tl) electromagnetic
calorimeter (ECL).

For this all-neutral final state, we require that there be no
reconstructed tracks coming from the vicinity of
the nominal collision point. 
Therefore, the CDC is used for vetoing events with charged track(s). 
Photons from decays of two neutral pions are detected and their 
momentum vectors are measured by the ECL. 
Signals from the ECL are also used to trigger signal events.

\section{Deriving differential cross sections}
\label{sec:cross}
The event triggers, data processing, 
and event selection are the same as those described in Ref.~\cite{pi0pi0}.
We derive the c.m. energy $W$ of the two-photon collision 
from the invariant mass of the two neutral pion system.
We calculate the cosine of the scattering angle of 
$\pi^0$ in the $\gamma \gamma$ c.m. frame, $| \cos \theta^*|$ 
for each event, using the $e^+e^-$ collision 
axis in the $e^+e^-$ c.m. frame as the reference of the 
polar angle as an approximation, because
we do not know the exact $\gamma\gamma$ collision axis. 

\subsection{Data reduction}
We find that the signal candidates in the low energy
region ($W<1.2~\GeV$) are considerably contaminated
by background events.  
We study the $p_t$-balance 
distribution, i.e., the event distribution in $|\sum \vec{p}_t^*|$,
to separate the signal and background components.
We estimate the $p_t$-unbalanced background component for $W<1.2~\GeV$,
and subtract the yield in the signal region, in the same manner as
in the previous analysis~\cite{pi0pi0}.  
However, above 1.2~GeV, we cannot quantitatively determine the background
contamination because of the small background rate and low statistics of the
sample, as well as the uncertainty in the functional form 
for the signal shape.

Using the ratio of yields between the $p_t$-balanced and 
unbalanced regions, we can estimate the backgrounds. 
In Fig.~1,
we plot the $W$ dependences of
\begin{equation}
R={\rm Yield}(0.15<|\sum \vec{p}_t^*| <0.20~\GeV/c)/
  {\rm Yield}(|\sum \vec{p}_t^*|<0.05~ \GeV/c).
\label{eqn:r_pt}
\end{equation}
We integrate over all angles in this figure.
The main part of the $W$-dependence of $R$ comes from
the energy dependence of the momentum resolution. 
The expected ratio from the pure signal component (calculated
from the signal Monte Carlo (MC) events and
corrected for the deviation of the $p_t$ resolution discussed in
the next subsection) is shown by the solid line. 
The excess of $R$ over the line ($\Delta R$) is expected to correspond
to the contribution from the $p_t$-unbalanced background.
The excess is relatively small above 1.0~GeV, although some fine structure
is visible there.
In the range 1.2-1.5~GeV, $\Delta R$ is undetectably small;
for 1.5-3.3~GeV, $\Delta R$ ranges between 0.00 and 0.08;
above $3.3~\GeV$, $\Delta R$ is the range from 0.08 to 0.2.
From the $R$ values, we estimate that the background 
contamination in the signal region is $\sim R/4$, which
is  smaller than 3\% for 1.1 - 3.3~GeV and around 3\% for 3.6 - 4.1~GeV. 
We subtract 3\% for 3.6 - 4.1~GeV, and assign a 3\% systematic 
error from this source for the 1.5 - 4.1~GeV range. 

We estimate the invariant-mass resolution from studies 
of signal-MC and experimental events. 
We find that an asymmetric Gaussian function with standard deviations of 
1.9\%$W$ and 1.3\%$W$ on the lower and higher sides of the peak, 
respectively, approximates the smearing reasonably well.  
Based on this information, unfolding is performed using the singular value 
decomposition (SVD) 
algorithm~\cite{svdunf} at the yield level~\cite{pi0pi0}, 
and is applied to obtain
the corrected $W$ distribution in the 0.9 - 2.4~GeV region,
using data with observed $W$ values between 0.72 and 2.68~GeV.
For lower energies, $W < 0.9~\GeV$, the effect of the migration is 
expected to be small because the invariant-mass resolution is small 
compared with the bin width. 
For higher energies, $W > 2.4~\GeV$, 
where the statistics is relatively low and unfolding would enlarge 
the errors, we rebin events into 100~MeV bins without  unfolding.

\begin{figure}
\centering
\includegraphics[width=9cm]{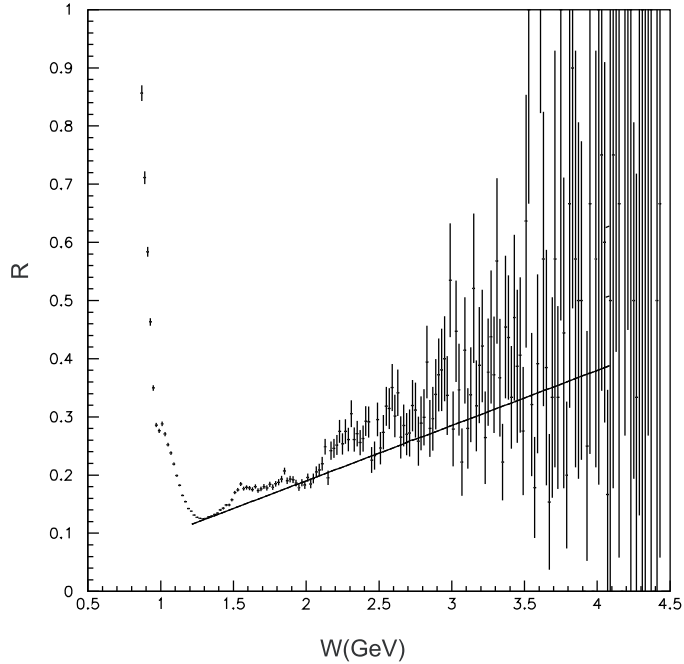}
\label{fig:fig1}
\centering
\caption{
The yield ratio $R$ in the $p_t$-unbalanced
bin to the $p_t$-balanced (signal) bin (see text for the
exact definition) for the experimental data.
The solid line shows the signal-component obtained
from the signal-MC and corrected taking into account the poorer momentum
resolution in experimental data.
}
\end{figure}

\subsection{Calculation of differential cross section}
We determine the trigger efficiency for
signal using the detector and trigger
simulators applied to the signal MC events. 
 The signal MC data for $e^+e^- \to e^+e^- \pi^0\pi^0$ are
generated using the TREPS code~\cite{treps} for the efficiency
determination, isotropically in $|\cos \theta^*|$ at 58 fixed $W$ points 
between 0.5 and 4.5~GeV,
The angular distribution at 
the generator level does not play a role in the efficiency
determination, because we calculate the efficiencies separately 
in each $|\cos \theta^*|$ bin with a 0.05 width.
Samples of $4 \times 10^5$ events are generated at each $W$ point. 
Two sets of different background conditions, which were extracted from
the beam collision data are embedded in the signal MC
data in the detector simulation; these are put through
the trigger simulator and the event selection program.
To minimize statistical fluctuations in the MC calculation, 
we fit the numerical results of the trigger efficiency to a
two-dimensional empirical function in $(W, |\cos \theta^*|)$.

The efficiency calculated from the signal MC events is 
corrected for a systematic difference of the peak 
widths in the $p_t$-balance distributions found between
the experimental data and the MC events, which is attributed
to a difference in the momentum resolution for $\pi^0$'s.
The correction factor is typically 0.95.

The DCS for each
($W$, $|\cos \theta^*|$) point is derived
from the following formula:
\begin{equation}
\frac{d\sigma}{d|\cos \theta^*|} =
\frac{\Delta Y - \Delta B}{\Delta W \Delta |\cos \theta^*| 
\int{\cal L}dt L_{\gamma\gamma}(W)  \eta } \; ,
\label{eqn:diffc}
\end{equation}
where $\Delta Y$ and $\Delta B$ are the signal yield and
the estimated $p_t$-unbalanced background in the bin, 
$\Delta W$ and $\Delta |\cos \theta^*|$ are the bin widths, 
$\int{\cal L}dt$ and  $L_{\gamma\gamma}(W)$ are
the integrated luminosity and two-photon luminosity function
calculated by TREPS~\cite{treps}, respectively, and  $\eta$ is the efficiency 
including the correction described above.
The energy bin width $\Delta W$ is 0.02~GeV for $0.6~\GeV < W < 1.8~\GeV$,
0.04~GeV for $1.8~\GeV < W < 2.4~\GeV$, and 0.1~GeV for 
$2.4~\GeV < W < 4.1~\GeV$. The width of each angular bin is
$\Delta |\cos \theta^*|=0.05$.

Figure~2
shows the $W$ dependence of the
cross section integrated over $|\cos \theta^*| < 0.8$. 
The results are obtained by simply summing
$d\sigma/d|\cos \theta^*| \cdot  \Delta |\cos \theta^*|$ 
over the corresponding angular bins. 
We have removed the bins in the range $3.3~\GeV < W < 3.6~\GeV$, 
because we cannot separate the $\chi_{c0}$ and $\chi_{c2}$  
components and the continuum in a model-independent way 
due to the finite mass resolution and insufficient statistics of 
the measurement. 
The cross section in this region is 
discussed in detail in Section~\ref{sec:highe}.

We show the angular dependence of the DCS
at several $W$ points in Fig.~3.
Note that the cross sections in the neighboring bins after the unfolding are 
no longer independent of each other
in both central values and size of errors.

\begin{figure}
\centering
\includegraphics[width=9cm]{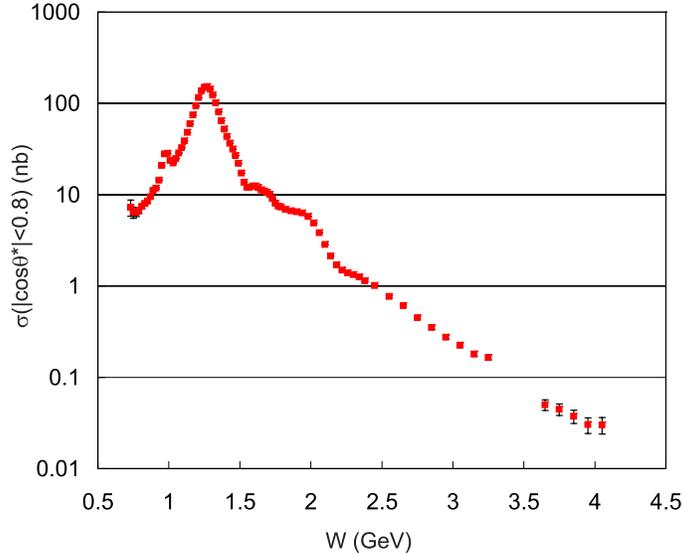}
\label{fig:fig2}
\centering
\caption{The integrated cross section results
in the angular regions $|\cos \theta^*|<0.8$.
Data points in bins near 3.5~GeV are not shown because of uncertainty
from the $\chi_{cJ}$ subtraction.}
\end{figure}

\begin{figure}
\centering
\includegraphics[width=14cm]{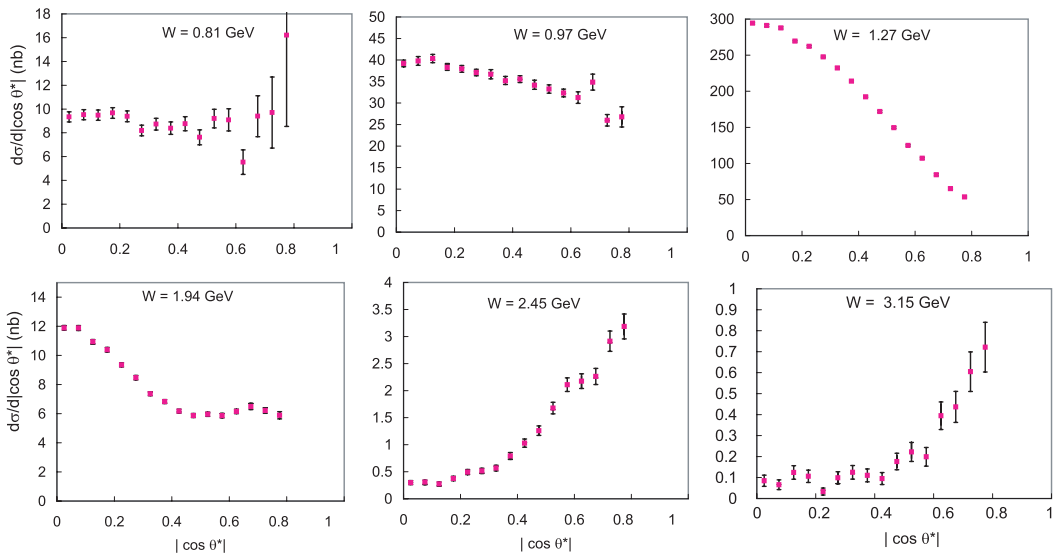}
\label{fig:fig3}
\centering
\caption{The DCS for five selected $W$ points, 0.97~GeV,
1.27~GeV, 1.95~GeV, 2.45~GeV and 3.15~GeV.}
\end{figure}

We estimate the systematic errors for the cross section in each 
energy bin arising from various sources. 
The systematic errors arise from uncertainties in
trigger efficiency (4-30\%), $\pi^0$ reconstruction efficiency (6\%),
$p_t$-balance cut (typically 3\%), background subtraction (0-40\%),
luminosity function (4-5\%), beam background effect for efficiency 
(2-4\%), other efficiency errors (4\%) and the unfolding procedure (0-4\%),
for which we show a range when the relative error size depends on $W$.
The total systematic error is obtained by adding the uncertainties
in quadrature and is 10\% in the intermediate $W$ region. 
The systematic error becomes much larger at lower $W$.
At higher $W$, the systematic error is rather
stable, typically $\sim 11$\%.

\section{Analysis of resonances in the range $1.4~\GeV < W < 2.2~\GeV$}
\label{sec:reson}
Previously, we have obtained a reasonable fit to a simple model
of resonances and smooth backgrounds in the energy region
$0.8~\GeV < W < 1.6~\GeV$ from the DCS
of $\gamma \gamma \rightarrow \pi^0 \pi^0$ with a 95~fb$^{-1}$
data sample~\cite{pi0pi0}.
The clear $f_0(980)$ peak and the large contribution from
the $f_2(1270)$ can be fitted with 
parameters determined from $\pi^+ \pi^-$ data~\cite{mori}.

In this section, we extend the analysis to the higher
energy region $1.4~\GeV < W < 2.2~\GeV$ (which has some overlap 
with the previous study) using a higher statistics sample of 
223~fb$^{-1}$.
It is well known that deriving reliable results on partial waves is 
difficult, especially in the higher energy region considered here.
The goal of this study is to provide some information on partial waves
in this energy region and to demonstrate the sensitivity (and limitations)
of our high-statistics data sample
(three orders of magnitude more statistics than the past experiments)
with a large angular coverage ($|\cos \theta^*| \le 0.8$).

\subsection{Parameterization of Partial Wave Amplitudes}
In the energy region $W \leq 3$~GeV, $J > 4$ partial waves (next is 
$J=6$) may be neglected so that only S, D and G waves are to be considered.
The DCS can be expressed as:
\begin{equation}
\frac{d \sigma}{d \Omega} (\gamma \gamma \to \pi^0 \pi^0)
 = \left| S \: Y^0_0 + D_0 \: Y^0_2  + G_0 \: Y^0_4 \right|^2 
+ \left| D_2 \: Y^2_2  + G_2 \: Y^2_4 \right|^2 \; ,
\label{eqn:diff}
\end{equation}
where $D_0$ and $G_0$ ($D_2$ and $G_2$) denotes the helicity 0 (2) components
of the D and G wave, respectively, and $Y^m_J$ are the spherical harmonics.

We derive some information on resonances in the
DCS fit by parameterizing partial wave amplitudes
in terms of resonances and smooth ``backgrounds''.
Once the functional forms of the amplitudes are assumed, we can use 
Eq.~(\ref{eqn:diff}) to fit the DCS.
From Fig.~10 in Ref.~\cite{pi0pi0} (also from the same plot with 
higher statistics), it appears that the 
G waves are non-zero at $W \simgt 1.8~\GeV$ and are dominated by 
the G$_2$ wave.
Here we assume (and check the necessity of) including
the $f_4(2050)$ in the G$_2$ wave; we note that the $f_4(2050)$'s
two-photon coupling has not been measured. 
Since the G$_2$ wave interferes with the D$_2$ wave, and the D$_2$
wave should contain resonances besides the $f_2(1270)$,
we include the known resonances, the  $f_2'(1525)$ and $f_2(1950)$,
which are known to couple to two photons 
(while the $f_2(2010)$ does not).
There are several more resonances that might couple to $\gamma \gamma$
and $\pi \pi$ in this mass region as listed in Ref.~\cite{pdg}.
Thus we assume that the $f_2(1950)$ is just an empirical 
parameterization representing these other resonances;
we denote it as the ``$f_2(1950)$''.

We parameterize the partial waves as follows:
\begin{eqnarray}
S &=& A_{f_0(Y)} e^{i \phi_{sY}} + B_S , \nonumber \\
D_0 &=& \sqrt{\frac{r_{02}}{1+r_{02}}} A_{f_2(1270)} e^{i \phi_{d0}} 
+ B_{D0}, \nonumber \\
D_2 &=& \sqrt{\frac{1}{1+r_{02}}} A_{f_2(1270)} e^{i \phi_{d2}} 
+ A_{f_2'(1525)} e^{i \phi_{2p}} + A_{f_2(Z)} e^{i \phi_{2Z}} + 
A_{``f_2(1950)''} e^{i \phi_{29}} + B_{D2}, \nonumber \\
G_0 &=& 0, \nonumber \\
G_2 &=& A_{f_4(2050)} e^{i \phi_{4}} + B_{G2}
\label{eqn:param2}
\end{eqnarray}
where $A_{f_0(Y)}$, $A_{f_2(1270)}$, $A_{f_2'(1525)}$, $A_{f_2(Z)}$, 
$A_{``f_2(1950)''}$ and $A_{f_4(2050)}$ are the
amplitudes of the corresponding resonances;
$B_S$, $B_{D0}$, $B_{D2}$ and $B_{G2}$ are ``background'' amplitudes for 
S, D$_0$, D$_2$ and G$_2$ waves; 
$r_{02}$ is the helicity-0 fraction of the $f_2(1270)$; 
and $\phi_{sY}$, $\phi_{d0}$, $\phi_{d2}$, $\phi_{2P}$, 
$\phi_{2Z}$, $\phi_{29}$ and $\phi_{4}$
are the phases of resonances relative to background amplitudes.
An $f_0(Y)$ term was needed to obtain a good fit in the previous analysis of
the energy region $0.8~\GeV < W < 1.6~\GeV$~\cite{pi0pi0}.
The $f_2(Z)$ is needed here to ``explain'' the dip-bump structure in
the range $1.5~\GeV < W < 1.7~\GeV$ as seen in Fig.~2.
We assume that $G_0 = 0$ and that $G_2$ consists only of the $f_4(2050)$
and a smooth ``background''.

We parameterize resonances with the formula given in Eq.~(\ref{eqn:arj}).
The relativistic Breit-Wigner resonance amplitude
$A_R(W)$ for a spin-$J$ resonance $R$ of mass $m_R$ is given by
\begin{eqnarray}
A_R^J(W) &=& \sqrt{\frac{8 \pi (2J+1) m_R}{W}} 
\frac{\sqrt{ \Gamma_{\gamma \gamma} \Gamma_{\pi^0 \pi^0}}}
{m_R^2 - W^2 - i m_R \Gamma_{\rm tot}} \; .
\label{eqn:arj}
\end{eqnarray}
Energy-dependent widths are used for
the parameterization of the $f_2(1270)$ and $f_2'(1525)$~\cite{mori}.
The resonance parameters given in Ref.~\cite{pdg} for the 
$f_2'(1525)$, $f_2(1950)$ and $f_4(2050)$
are summarized in Table~\ref{tab:param}.
Since all (some) of the individual decay fractions of the ``$f_2(1950)$'' 
($f_4(2050)$) are not known, we neglect the $W$ dependence of their partial
and total widths.
\begin{center}
\begin{table}
\caption{Parameters of the $f_2'(1525)$, $f_2(1950)$ and 
$f_4(2050)$~\cite{pdg}.}
\label{tab:param}
\begin{tabular}{lcccc} \hline \hline
Parameter  & $f_2'(1525)$  & $f_2(1950)$  & $f_4(2050)$ & Unit \\
\hline
Mass & $1525 \pm 5$ & $1944 \pm 12$ & $2025 \pm 10$ & MeV/$c^2$ \\
Width & $73^{+6}_{-5}$ & $472 \pm 18$ & $225 \pm 18$ & MeV \\
${\cal B} (\pi \pi)$ & $0.82 \pm 0.15$ & seen & $17.0 \pm 1.5$ & \% \\
${\cal B} (K \bar{K})$ & $88.8 \pm 3.1$ & seen & $0.68^{+0.34}_{-0.18}$
& \% \\
${\cal B} (\eta \eta)$ & $10.3 \pm 3.1$ & seen & $21 \pm 8$& \% \\
${\cal B} (\gamma \gamma)$  & $1.11 \pm 0.14$ & seen & unknown & $10^{-6}$\\
\hline\hline
\end{tabular}
\end{table}
\end{center}


Background amplitudes are parameterized as follows.
\begin{eqnarray}
B_S &=&  a_{sr} (W - W_0)^2  + b_{sr} (W - W_0) + c_{sr}
 + i (a_{si} (W - W_0)^2  + b_{si} (W - W_0)  + c_{si}) , \nonumber \\
B_{D0} &=& a_0 (W - W_0)^2 + b_0 (W - W_0) + c_0 , \nonumber \\
B_{D2} &=& a_2 (W - W_0)^2 + b_2 (W - W_0) + c_2 , \nonumber \\
B_{G2} &=&  a_{gr} (W - W_0)^2  + b_{gr} (W - W_0) + c_{gr}
 + i (a_{gi} (W - W_0)^2  + b_{gi} (W - W_0)  + c_{gi}) ,\nonumber \\
\label{eqn:para4}
\end{eqnarray}
where $W_0 = 1.2$~GeV and we fix $c_{gr} = c_{gi} = 0$ to reduce the number
of parameters; floating them does not improve fitting.
We assume background amplitudes to be quadratic
in $W$ for all the waves for both real and imaginary parts.
The background $D_0$ and $D_2$
amplitudes are taken to be real by definition.

\subsection{Fit results}
In the fit, we fix the values of the parameters for the $f_0(Y)$,
$f_2(1270)$ and the phases $\phi_{d0}$ and $\phi_{d2}$ to those
determined in Ref.~\cite{mori} and \cite{pi0pi0}.
The parameters of the $f_2'(1525)$ are fixed to those in Ref.~\cite{pdg}.
If the resonance parameters of the ``$f_2(1950)$'' and $f_4(2050)$
are fixed at the values given in
PDG~\cite{pdg} as summarized in Table~\ref{tab:param}, then
the fit is very poor yielding $\chi^2/ndf = 4.0$.
Here we quote the results when they are floated.
By fixing the phases $\phi_{d0}$ and $\phi_{d2}$, the sign ambiguity 
for $B_{D0}$ and $B_{D2}$ can be resolved.

Here the unfolded DCS are fitted.
One to three thousand sets of randomly generated initial parameters are 
fitted for each study.
Two solutions (denoted by sol.~A and B)
of reasonably good fit quality ($\chi^2/ndf = 1.08$)
are obtained only for the nominal fit, where an $f_2(Z)$ component
is included together with the ``$f_2(1950)$'' and $f_4(2050)$, whose
parameters are floated. 
Fit results are shown in Fig.~4
for the DCS and in Fig.~5
for the total cross section.
Two solutions are indistinguishable.
The main difference between sol.~A and B is the two photon coupling of
the ``$f_2(1950)$'' and $f_2(Z)$; that is 
larger by a factor of 8 (5) in sol.~B for the ``$f_2(1950)$'' ($f_2(Z)$),
which may be too large.
These two solutions arise from constructive and destructive interference
between resonances and backgrounds.

Since the two-photon coupling of the $f_4(2050)$ has not been measured,
a fit without it is also given in Table~\ref{tab:fit1}.
The fit quality is unacceptable, indicating that the $f_4(2050)$ has a
non-zero two-photon coupling.
The existence of the $f_2(Z)$, a spin-2 resonance with mass near 
1500~MeV$/c^2$, is controversial.
Thus, fits are also made without the $f_2(Z)$.
There, two fits are carried out fixing and floating the branching fraction to
two photons of the $f_2'(1525)$.
The fitted parameters are listed in Table~\ref{tab:fit1}.
The fit without the $f_2(Z)$ gives an unacceptable $\chi^2$.
When the branching fraction of the $f_2'(1525)$ to two photons 
is floated with no $f_2(Z)$ contribution, the value obtained is 
ten times larger than the nominal value,
which is unacceptable; thus some resonance like the $f_2(Z)$ is indeed
necessary.

\begin{figure}
 \centering
\includegraphics[width=8cm]{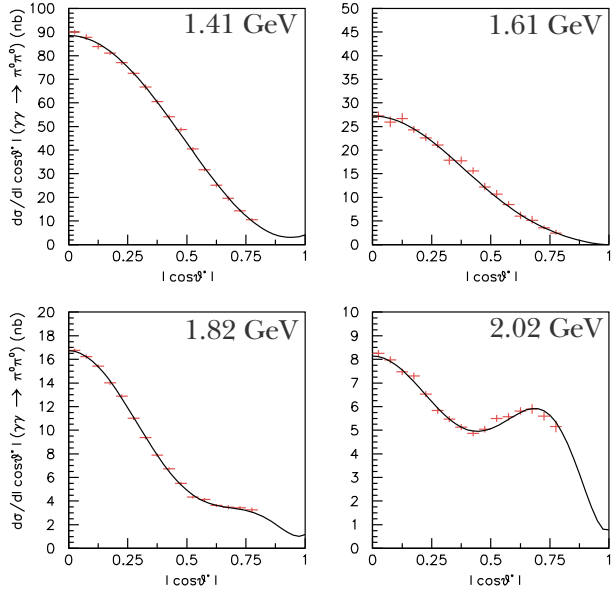}
 \caption{Fitted curves (solid line) and DCS
($d \sigma / d |\cos \theta^*|$ (nb))
for $W$-bins indicated.}
\label{fig:fig4}
\end{figure}
\begin{figure}
 \centering
\includegraphics[width=8cm]{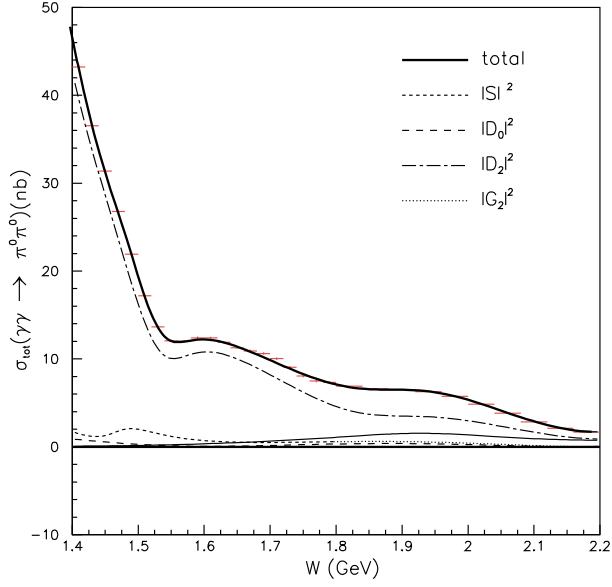}
 \caption{Total cross section 
($|\cos \theta^*| < 0.8$ (nb)) and fitted curves.}
\label{fig:fig5}
\end{figure}


\begin{center}
\begin{table}
\caption{Fitted parameters (1)} 
\label{tab:fit1}
\begin{tabular}{lcccccc} \hline \hline
Parameter &\multicolumn{3}{c}{With  $f_2(Z)$} 
& \multicolumn{2}{c}{Without $f_2(Z)$} & \\
& \multicolumn{2}{c}{All included} & no 
& \multicolumn{2}{c}{${\cal B} (f_2'(1525) \rightarrow \gamma \gamma )$} 
& Unit\\
& sol.~A & sol.~B & $f_4(2050)$ & fixed & free & \\
\hline
Mass$(f_4(2050))$ &  $1935^{+12}_{-14}$ &  $1902^{+12}_{-13}$ & --
& $1865^{+14}_{-15}$ & $1866 \pm 19$ & MeV/$c^2$ \\
$\Gamma_{\rm tot}(f_4(2050))$ & $369^{+17}_{-22}$  & $350^{+25}_{-22}$ & --
& $379^{+34}_{-29}$ & $356^{+32}_{-29}$ & MeV\\
$\Gamma_{\gamma \gamma} (f_4(2050))$ &  $45^{+11}_{-19}$ &  $74^{+16}_{-13}$ 
& 0 (fixed) & $91^{+32}_{-25}$ & $68 \pm 12$ & eV\\
\hline
mass$(``f_2(1950)'')$ & $1852^{+23}_{-20}$  & $1906^{+8}_{-9}$  
&  $2191 \pm 6 $  & $1638^{+8}_{-7}$ & $1741^{+9}_{-12}$ & MeV/$c^2$\\
$\Gamma_{\rm tot}(``f_2(1950)'')$ & $347^{+23}_{-20}$ & $394^{+24}_{-22}$ 
&  $38^{+18}_{-14}$ & $357^{+16}_{-17}$ & $370^{+11}_{-12}$ & MeV\\
$\Gamma_{\gamma \gamma} {\cal B}(\pi^0 \pi^0)$
&  $9.2^{+4.8}_{-2.6}$  &  $75^{+24}_{-22}$ & $0.4^{+0.3}_{-0.2}$  
& $165.9^{+22.2}_{-24.1}$ & $50.3^{+54.4}_{-17.0}$ & eV \\
\hline
mass$(f_2(Z))$ &  $1526^{+9}_{-6}$ &  $1542^{+7}_{-5}$ 
&  $1649 \pm 5 $  & --& -- & MeV/$c^2$ \\
$\Gamma_{\rm tot}(f_2(Z))$ &  $121 \pm 9$  &  $217^{+15}_{-16}$ 
&  $414 \pm 15 $  & -- & -- & MeV \\
$\Gamma_{\gamma \gamma} {\cal B}(\pi^0 \pi^0)$ 
&  $17.5^{+2.8}_{-5.6}$  &  $87 \pm 27$  
&  $503^{+47}_{-40}$  & 0 (fixed) & 0 (fixed) & eV \\
\hline
${\cal B}_{f_2'(1525)} (\gamma \gamma )$ & 
\multicolumn{4}{c}{1.11 (fixed)}
& $11.9 \pm 1.3$ & $10^{-6}$\\ \hline
\hline
$\chi^2 \; (ndf)$ & 485.0 (450) & 485.2 (450) & 619.1 (454) 
&  571.0 (454) & 517.5 (453) & \\
\hline\hline
\end{tabular}
\end{table}
\end{center}

The spin of the ``resonance'' of mass $\sim 1500$~MeV$/c^2$ (denoted here as
$f_2(Z)$ for a spin-2 resonance) is not known.
Thus we also fit by assuming the spin to be 0 denoted as the $f_0(Z)$,
which contributes to the S-wave.
In additions, fits with the $f_4(2050)$ only, the ``$f_2(1950)$'' 
only and no new resonances are performed;
the results are summarized in Table~\ref{tab:fit2}.
The $f_2(Z)$ hypothesis is favored over that of the $f_0(Z)$ with
about 3$\sigma$ significance, which is calculated from the difference of 
$\chi^2$ values.

\begin{center}
\begin{table}
\caption{Fitted parameters (2)} 
\label{tab:fit2}
\begin{tabular}{lccccc} \hline \hline
Parameter &With $f_0(Z)$ & $f_4(2050)$ & ``$f_2(1950)$'' 
& None & Unit \\ 
&& only & only &&\\ \hline
Mass$(f_4(2050))$  &  $1876^{+11}_{-10}$ 
& $1894 \pm 7$  & -- & -- & MeV/$c^2$\\
$\Gamma_{\rm tot}(f_4(2050))$ &  $493^{+17}_{-20}$  
& $268^{+15}_{-13}$ & -- & -- & MeV \\
$\Gamma_{\gamma \gamma} (f_4(2050))$ &  $213^{+42}_{-43}$  
& $31^{+8}_{-5}$ & & 0 (fixed)  & eV \\
\hline
Mass$(``f_2(1950)'')$  &  $1752^{+14}_{-9}$  & -- 
& $1630 \pm 31$ & -- & MeV/$c^2$\\
$\Gamma_{\rm tot}(``f_2(1950)'')$ & $310^{+26}_{-24}$  & -- 
& $362^{+11}_{-9}$ & -- & MeV \\
$\Gamma_{\gamma \gamma} {\cal B}(``f_2(1950)'' \to \pi^0 \pi^0)$ 
&  $10.0^{+4.9}_{-3.2}$  & 0 (fixed) & $132^{+12}_{-11}$ & 0 (fixed)  & eV \\
\hline
Mass$(f_0(Z))$  &  $1566^{+10}_{-13}$ & --& --& -- & MeV/$c^2$\\
$\Gamma_{\rm tot}(f_0(Z))$ &  $118^{+23}_{-27}$  & -- & -- & -- & MeV \\
$\Gamma_{\gamma \gamma} {\cal B}(f_0(Z) \to \pi^0 \pi^0)$ 
&  $91^{+52}_{-39}$  & 0 (fixed) & 0 (fixed) & 0 (fixed) & eV \\
\hline
$\chi^2 \; (ndf)$ & 496.3 (450) &  1938.6 (458) & 705.0 (458) 
& 2950.0 (462) & \\
\hline\hline
\end{tabular}
\end{table}
\end{center}

A study of systematic errors is not performed because
we do not know how to estimate uncertainty from model dependence.

\section{Analysis of the higher-energy region}
\label{sec:highe}
In general, we expect that theoretical models based on QCD 
give reasonable predictions even for two-photon production
of exclusive final-states
such as $\gamma \gamma \to \pi^0 \pi^0$
in the high energy region. 
However, the models do not give information on what energies can be considered
as high enough. 

The handbag model~\cite{handbag} predicts that the angular dependence 
of the DCS for $\gamma \gamma \to \pi\pi$ goes as 
$\sim \sin^{-4} \theta^*$. 
This prediction is common to charged and neutral pairs of pions. 
Our measurement for the charged-pion process agrees with 
this expectation above $W>3.1~\GeV$~\cite{nkzw}. 
The prediction of the cross section ratio,
$\sigma(\pi^0\pi^0)/\sigma(\pi^+\pi^-)$ is determined by isospin invariance
to be 0.5.

However, predictions based on perturbative QCD at leading order
suggests that an angular distribution for $\pi^0 \pi^0$ will be different 
from  that for $\pi^+ \pi^-$.
The primary term of the leading order (a short-range emission term)
for the charged-pion process is
well described by a ~$\sin^{-4} \theta^*$ dependence~\cite{bl},
while this term vanishes for the neutral-pion process.
The shape of the angular distribution from the next 
term (a long-range interaction term)
is unpredictable, and in general, perturbative QCD models predict smaller
cross sections for this next term and hence a small neutral to
charged cross-section ratio, $\sigma(\pi^0\pi^0)/\sigma(\pi^+\pi^-)\sim 0.03$.

\subsection{Angular dependence}
We compare the angular dependence of the DCS in the range
$|\cos \theta^*|<0.8$ for
$W > 2.4~\GeV$ with the function $\sin^{-4} \theta^*$.
We also try a fit with an additional term, 
to quantify a possible deviation from  $\sin^{-4} \theta^*$ behavior.
We choose this function because it gives relatively good fits 
empirically in a wide range of $W$.
Thus the fit function is parameterized as:
\begin{equation}
d\sigma/d|\cos \theta^*| = a(\sin^{-4} \theta^* + b\cos^2 \theta^*) .
\label{eqn:angul}
\end{equation}
We fit using a binned maximum likelihood method and 16 bins
in  the range $|\cos \theta^*|<0.8$.  
We know that the effect of charmonia is large in the region 
$3.2~\GeV < W < 3.6~\GeV$, but we cannot separate it
in the angular dependence because we cannot assume here any functional 
shapes for the non-charmonium component. 
The results of the fit for $b$ are shown in Fig.~6
as well as the fit to the angular distributions in
the four selected $W$ regions, 
where the DCS, the vertical axis of this figure, is normalized to the total
cross section $\sigma(|\cos \theta^*|<0.8)$ in each $W$ region, i.e. 
the area under the curve is 1.
The parameter $b$ is close to zero above $W > 3.1~\GeV$
(comparing with $b=9.279$, which would give the same contribution
in $\sigma(|\cos \theta^*|<0.8)$ as the $\sin^{-4} \theta^*$ term does)
although
it becomes nearly constant and then systematically negative above 
the charmonium region.
The change in the $b$ parameter, which approaches a constant value near zero,
occurs at a $W$ value close to that observed in the charged
pion case.

\begin{figure}
\centering
\includegraphics[width=8cm]{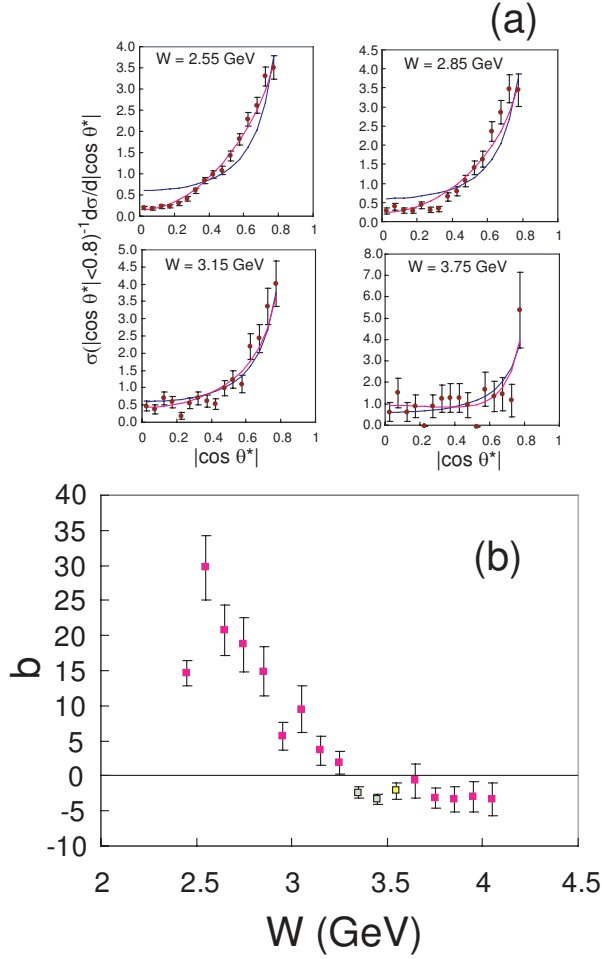}
\label{fig:fig6}
\centering
\caption{(a) The fits of the angular dependence of the normalized DCS 
(see text) at four selected $W$ points.
For the blue curves the coefficient $b$ (see the fit formula in the text) 
is fixed to 0. 
The magenta curves show the fits with $b$ floating. 
(b) The energy dependence of the parameter $b$ giving the best fits. 
Here, the charmonium contributions are not subtracted, and
the data in the $\chi_{c0}$ and $\chi_{c2}$ charmonium 
regions are plotted with different colors.}
\end{figure}

\subsection{Yields of $\chi_{cJ}$ charmonia}
The structures seen in the yield distribution 
for $3.3~\GeV < W < 3.6~\GeV$ (Fig.~7)
are from charmonium production, 
$\gamma \gamma \to \chi_{c0}$, $\chi_{c2} 
\to \pi^0\pi^0$. 
Similar production of the two charmonium states
is observed in the $\pi^+\pi^-$, $K^+K^-$ and $K^0_SK^0_S$ final 
states~\cite{nkzw,wtchen}.
 
We fit the distribution to contributions of $\chi_{c0}$,
$\chi_{c2}$ and a smooth continuum component,using the following function: 
\begin{equation}
Y(W) = |\sqrt{\alpha kW^{-\beta}}+e^{i\phi}\sqrt{N_{\chi_{c0}}}
{\rm BW}_{\chi_{c0}}(W)|^2 + 
N_{\chi_{c2}}|{\rm BW}_{\chi_{c2}}(W)|^2 + \alpha (1-k)W^{-\beta},
\label{eqn:chicj}
\end{equation}
in the $W$ region between 2.8 and 4.0~GeV,
where ${\rm BW}_{\chi_{cJ}}(W)$ is a Breit-Wigner function for the charmonium
amplitude, which is proportional to 
$\sim 1/(W^2-M_{\chi_{cJ}}^2-iM_{\chi_{cJ}} \Gamma_{\chi_{cJ}})$
and is normalized as $\int |{\rm BW}_{\chi_{cJ}}(W)|^2dW=1$.
The masses and widths, $M$ and $\Gamma$, of the charmonium states
are fixed to the PDG world averages~\cite{pdg}.
The component $\alpha W^{-\beta}$ corresponds to the contribution from 
the continuum component, 
with a fraction $k$ that interferes with the $\chi_{c0}$ amplitude 
with a relative phase angle, $\phi$.  
It is impossible to determine the interference parameters for the
 $\chi_{c2}$, because of its much smaller intrinsic width compared 
to the measurement resolution. 
We fit the $\chi_{c2}$ yield ($N_{\chi c2}$) with a formula where
no interference term is included, and later we estimate the maximum
effects from the interference term when deriving the two-photon
decay width of $\chi_{c2}$.
We use data only in the range $|\cos \theta^*|<0.4$ where the charmonium 
contribution is dominant. 
We take into account the smearing effect 
due to a finite mass resolution in the fit, using the same function
as used for the unfolding.

A binned maximum likelihood method is applied. 
We examined two cases with and without the interference. 
Reasonably good fits are obtained for both cases.
The fit results are summarized in Table~\ref{tab:charm1}. 
In the table, ${\cal L}$ is the likelihood value
and $ndf$ is the number of degrees of freedom.
The normalization $N_{\chi c0}$ in Eq.(8) is proportional to 
the square of the resonance amplitude.
The yields from the fits are translated into products 
of the two-photon
decay width and the branching fraction, 
$\Gamma_{\gamma \gamma}(\chi_{cJ}){\cal B}(\chi_{cJ} \to \pi^0\pi^0)$, 
which are listed in Table~\ref{tab:charm2}. 
The systematic errors are taken from the changes in the best fits
in the central values of yields when the absolute energy
scale is varied by $\pm 10$~MeV for the $W$ measurement and the
invariant-mass resolution is varied by $\pm 10$\% for the corresponding
Gaussian widths. 
The changes of the goodness of fit ($-2\ln{\cal L}$) for these variations
are found to be small, less than 1.7.

The $\chi_{c0}$ is observed with a statistical significance of $7.6\sigma$
($7.3\sigma$) when we take (do not take) interference into account.
The statistical significance for the $\chi_{c2}$ is $2.6\sigma$ when
we take interference of the $\chi_{c0}$ into account, but it is
only $1.3\sigma$ when we do not take into account interference. 
This is because interference makes the line shape
of $\chi_{c0}$ highly asymmetric with a short tail and destructive 
interference on the high-energy side.
The red and blue curves in Fig.~7
show the fits for the two cases
of with and without $\chi_{c0}$ interference.

The results for $\Gamma_{\gamma\gamma}{\cal B}(\chi_{cJ})$ 
in the $\pi^0\pi^0$ final state are compared with previous
measurements of  $\Gamma_{\gamma\gamma}{\cal B}(\chi_{cJ})$ in
the $\pi^+\pi^-$ decay mode,
$15.1 \pm 2.1 \pm 2.3$~eV and $0.76 \pm 0.14 \pm 0.11$~eV for
$\chi_{c0}$ and $\chi_{c2}$, respectively~\cite{nkzw},
(or even $K\bar{K}$, referring to SU(3) symmetry~\cite{nkzw,wtchen})
decay mode. 
Although the effects of interference were neglected
in the $\pi^+\pi^-$ measurements, the results are consistent
with the ratio expected from isospin invariance, 
${\cal B}(\chi_{cJ} \to \pi^0\pi^0)/{\cal B}(\chi_{cJ} \to \pi^+\pi^-)=1:2$.

\small
\begin{center}
\begin{table}
\caption{Results of the fits (see text) to obtain the charmonium
contributions with and without interference effects. 
Errors are statistical only. Logarithmic likelihood
($\ln {\cal L}$) values are only meaningful when comparing two or more
fits.}
\label{tab:charm1}
\begin{tabular}{c|ccccc}
\hline
Interference & $N_{\chi_{c0}}$ & $k$ & $\phi$ &
$N_{\chi  c2}$ & $-2\ln{\cal L}/ndf$ \\
\hline
Without & $100 \pm 16$ & $-$ & $-$ & $13^{+11}_{-10}$ & $52.4/56$\\
With & $103^{+60}_{-42}$ & $0.82^{+0.18}_{-0.48}$ & $(1.1 \pm 0.3)\pi$ 
& $34 \pm 13$ & $44.2/54$\\
\hline
\end{tabular}
\end{table}
\end{center} 
\ \\
\begin{center}
\small
\begin{table}
\caption{Products of the two-photon decay width
and the branching fraction for the two charmonia.
Here,  $\Gamma_{\gamma \gamma}{\cal B}(\chi_{cJ})$
means $\Gamma_{\gamma \gamma}(\chi_{cJ}){\cal B}(\chi_{cJ} \to \pi^0\pi^0)$.
The first, second and third errors (if exists) are statistical, systematic
and from the maximal uncertainties of the relative phase in $\chi_{c2}$
production.}
\label{tab:charm2}
\begin{tabular}{c|cc}
\hline
Interference & $\Gamma_{\gamma \gamma}{\cal B}(\chi_{c0})$ (eV)& 
$\Gamma_{\gamma \gamma}{\cal B}(\chi_{c2})$ (eV)\\
\hline
Without & $9.7 \pm 1.5 \pm 1.0$ & $0.18^{+0.15}_{-0.14} \pm 0.03$ \\
With & $9.9^{+5.8}_{-4.0} \pm 1.0$ & $0.48 \pm 0.18 \pm 0.05 \pm 0.14$ \\
\hline
\end{tabular}
\end{table}
\end{center} 
  
\begin{figure}
\centering
\includegraphics[width=10cm]{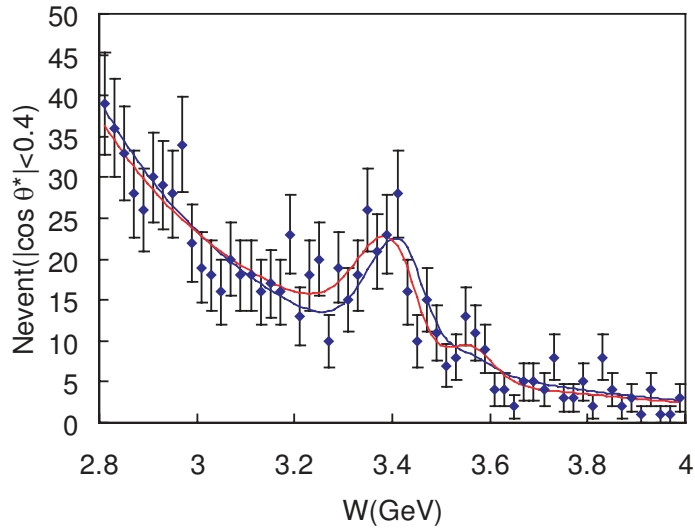}
\label{fig:fig7}
\centering
\caption{The $W$ distribution
of the candidate events with $|\cos \theta^*|<0.4$  
near the charmonium region.
The red and blue curves show the fits described in the text with and
without interference with the $\chi_{c0}$.}
\end{figure}
\normalsize

\subsection{Subtraction of the charmonium contributions}
We subtract the charmonium contributions from nearby bins
of the charmonium ($\chi_{cJ}$) region, $3.2 - 3.6~\GeV$,
in order to obtain a pure DCS
from the continuum component.
We use the fit result with interference obtained in the previous subsection. 

The estimated charmonium yield that includes the contribution from the
interference term is converted to a DCS contribution in each angular
bin of $|\cos \theta^*|<0.8$ 
by assuming a flat distribution 
for the $\chi_{c0}$ component and a
$\sim \sin^4 \theta^*$ distribution
for the $\chi_{c2}$ component~\cite{wtchen}.
This assumption is only a model. 
In reality, we do not know how the angular distribution
of the interference term behaves; the charmonium amplitudes 
can interfere with the continuum 
components with different $J$'s of unknown sizes.

 For the $W=3.25$~GeV bin, the fit result indicates that there is
a non-negligible effect from $\chi_{c0}$ when we assume interference,
and thus we make a correction for charmonium subtraction.  
The contribution of the charmonium components in the original DCS is 
18\% at $|\cos \theta^*|<0.6$. 
For  $W=3.3 - 3.6$~GeV,  we apply subtraction  
for the angular bins  $0.4<|\cos \theta^*| <0.8$ after extrapolating 
the charmonium yield determined in the range $|\cos \theta^*| <0.4$.

 The DCS thus obtained for the continuum is integrated over the range
$|\cos \theta^*|<0.6$.
We convert $\sigma(0.4<|\cos \theta^*| <0.8)$ to  
$\sigma(|\cos \theta^*| <0.4)$ for $W=3.3 - 3.6$~GeV, 
by assuming that the angular dependence of the
DCS has a $\sim \sin^{-4} \theta^*$ dependence. 
The results are plotted in Fig.~8.

\subsection{$W$ dependence and cross-section ratio}
 We fit the DCS integrated over angle, $\sigma(|\cos \theta^*|<0.6)$, 
to a power law in the c.m. energy, $W^{-n}$, for the energy region 
$3.1~\GeV <W< 4.1~\GeV$, in which the angular
dependence of the DCS does not show any large changes. 
In the fit, we do not use the data in the
charmonium region ($W=3.3 - 3.6$~GeV), where we cannot determine 
the cross section
of the continuum component in a model-independent manner.

 The obtained result is $n=6.9 \pm 0.6 \pm 0.7$. 
The systematic error is dominated by the uncertainty of the charmonium
contribution in $3.1~\GeV < W < 3.3~\GeV$.
This value is compatible with 
the results for the $\pi^+\pi^-$ and $K^+K^-$ processes~\cite{nkzw}, 
but significantly different
from the case of  $K^0_SK^0_S$~\cite{wtchen}. 

\begin{figure}
\centering
\includegraphics[width=8cm]{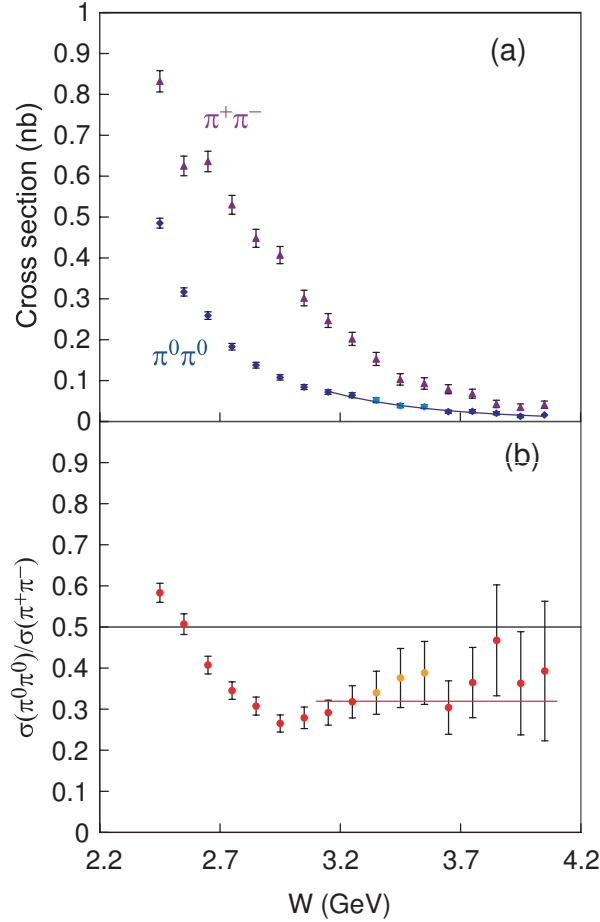}
\label{fig:fig8}
\centering
\caption{(a) The cross sections for the
$\gamma\gamma \to \pi^0\pi^0$ (blue diamonds) and
$\gamma \gamma \to \pi^+\pi^-$ (violet triangles, \cite{nkzw})
for $|\cos \theta^*|<0.6$.
The curve is the fit
to the cross section for $\gamma\gamma \to \pi^0\pi^0$ 
with a $\sim W^{-n}$ functional shape.
(b) Ratio of the cross section of the $\pi^0\pi^0$ process
to the $\pi^+\pi^-$ process. 
The error bars are statistical only. 
The red line is the average for $3.1~\GeV < W < 4.1~\GeV$.
The horizontal line (0.5) is an expectation
from isospin invariance for a pure $I=0$ component.
In (a) and (b), the estimated charmonium contributions 
are subtracted in both $\pi^+\pi^-$ and $\pi^0\pi^0$ measurements.
The results in the $W$ region 3.3~GeV - 3.6~GeV (plotted 
with lighter colors) are not used for the fits.}
\end{figure}

The fit for  $3.1~\GeV < W < 4.1~\GeV$ is shown in Fig.~8(a).
We also show the ratio of $\pi^0\pi^0$
to $\pi^+\pi^-$  cross sections in Fig.~8(b).
The two processes have a similar $W^{-n}$ dependence for
$3.1~\GeV < W < 4.1~\GeV$,
while their ratio is almost constant in this energy region. 
The average of the ratio in this energy region is 
$<{\rm Ratio}>= 0.32 \pm 0.03 \pm 0.05$, where the data in the 
3.3 - 3.6~GeV region is not used when calculating this average.
This ratio is significantly larger than the prediction of the leading order
QCD calculations~\cite{bl,bc}
and is slightly smaller than the value of 0.5, which appears
in Ref.~\cite{handbag} based on isospin invariance.

\section{Summary and Conclusion}
\label{sec:concl}
We have measured the process $\gamma \gamma \to \pi^0\pi^0$ based on
data from $e^+e^-$ collisions corresponding to an integrated luminosity
of 223~fb$^{-1}$ with the Belle detector at the KEKB accelerator. 
We derive results for the differential cross sections
in the center-of-mass energy and polar angle ranges, 
$0.6~\GeV < W < 4.1~\GeV$ and $|\cos \theta^*|<0.8$. 

A simple model is employed to obtain some information on resonances up to
2.2~GeV/$c^2$.
Differential cross sections are fitted in the energy region, 
$1.4~\GeV < W <  2.2~\GeV$, with a model where partial waves consist of 
resonances and smooth backgrounds.
The G wave is seen to be important for $W > 1.8~\GeV$,
where the G$_2$ wave appears to dominate over the G$_0$ wave~\cite{pi0pi0}.
Thus the $f_4(2050)$ is included in the G$_2$ wave while G$_0$
is set to zero.
Parameters of the $f_2(1270)$ and other resonances are fixed at 
PDG~\cite{pdg} values
and those determined in the previous analysis~\cite{mori, pi0pi0}. 
Two solutions with reasonably good fit quality are obtained 
when an additional spin-2 resonance is introduced with 
a mass near 1.5~GeV/$c^2$ (denoted as $f_2(Z)$) in addition to 
the $f_2'(1525)$, $f_2(1950)$ and $f_4(2050)$.
Without the $f_4(2050)$, the fit quality is unacceptable.

We observe production of the charmonium $\chi_{c0}$ state and obtain the
product of its two-photon decay width and the branching fraction to 
$\pi^0\pi^0$.
The energy and angular dependences above 3.1~GeV are compatible 
with those measured in the $\pi^+\pi^-$ channel, and we obtain the 
cross section ratio,
$\sigma(\pi^0\pi^0)/\sigma(\pi^+\pi^+)$, to be $0.32 \pm 0.03 \pm 0.05$ 
on average in the 3.1-4.1~GeV region. 
This ratio is significantly larger than the prediction of the leading order
QCD calculation.

\section*{Acknowledgment}
We thank the KEKB group for the excellent operation of the
accelerator, the KEK cryogenics group for the efficient
operation of the solenoid, and the KEK computer group and
the National Institute of Informatics for valuable computing
and SINET3 network support. We acknowledge support from
the Ministry of Education, Culture, Sports, Science, and
Technology of Japan and the Japan Society for the Promotion
of Science; the Australian Research Council and the
Australian Department of Education, Science and Training;
the National Natural Science Foundation of China under
contract No.~10575109 and 10775142; the Department of
Science and Technology of India; 
the BK21 program of the Ministry of Education of Korea, 
the CHEP src program and Basic Research program (grant 
No. R01-2005-000-10089-0, R01-2008-000-10477-0) of the 
Korea Science and Engineering Foundation;
the Polish State Committee for Scientific Research; 
the Ministry of Education and Science of the Russian
Federation and the Russian Federal Agency for Atomic Energy;
the Slovenian Research Agency;  the Swiss
National Science Foundation; the National Science Council
and the Ministry of Education of Taiwan; and the U.S.\
Department of Energy.

\end{document}